\documentclass[twocolumn,showpacs,preprintnumbers,amsmath,amssymb]{revtex4}

\usepackage{graphicx}
\usepackage{dcolumn}
\usepackage{bm}
\usepackage{amsmath}

\begin{document}


\title{Evidence for a $\beta$-decaying 1/2$^-$ isomer in $^{71}$Ni}

\author{I. Stefanescu$^{1,2,3,4}$, D. Pauwels$^1$, N. Bree$^1$, T.E. Cocolios$^1$, J. Diriken$^1$, S. Franchoo$^5$, M. Huyse$^1$, O. Ivanov$^1$, Y. Kudryavtsev$^1$, N. Patronis$^1$, J. Van De Walle$^1$, P. Van Duppen$^1$, W.B. Walters$^2$}
 \affiliation{$^1$Instituut voor
Kern- en Stralingsfysica, K.U. Leuven, Celestijnenlaan
200D, B-3001 Leuven, Belgium}
\affiliation{$^2$Department of Chemistry and Biochemistry, University of Maryland, College Park, Maryland 20742, USA}
\affiliation{$^3$Physics Division, Argonne National Laboratory, Argonne, Illinois 60439, USA}
\affiliation{$^4$Horia-Hulubei
National Institute for Physics and Nuclear Engineering, PO-Box
MG-6, Bucharest, Romania}
\affiliation{$^5$IPN Orsay, F-91406 Orsay Cedex France}

\date{\today}

\begin{abstract}

We report on the investigation of the population mechanism for the
454-KeV level in $^{71}$Cu. This level was identified for the
first time in a recent Coulomb excitation measurement with a
radioactive beam of $^{71}$Cu. The selective nature of the Coulomb-excitation process as well as nuclear-structure considerations
constrain the possible spin values for the newly observed state to
$I^\pi$=1/2$^-$. A re-examination of the data set
obtained in a $\beta$-decay study at the LISOL separator revealed that this state is also populated in the decay of $^{71}$Ni, most probably by direct
feeding from a newly identified 1/2$^-$ $\beta$-decaying isomer having a $T_{1/2}$=2.34(25) s. In this paper we
investigate the proposed scenario by reanalyzing the
$\beta$-$\gamma$ and $\gamma$-$\gamma$ coincidences obtained in the $\beta$-decay study at LISOL.

\end{abstract}

\pacs{25.70.De,21.10.Ky,21.60.Cs,27.50.+e}
\maketitle

\section{\bf INTRODUCTION}

Isomeric states in nuclei around closed shells can occur due to
the large spin difference between the valence orbitals
(single-particle isomers) or when the maximum amount of angular
momentum created in a single-particle configuration is reached
(seniority isomers). Depending on the transition probabilities involved, an isomeric state can decay by e.g., $\gamma$ transitions to lower lying-states or $\beta$ radiation to the ground or excited states of the daughter nucleus. The investigation of nuclear isomers gives important information regarding the evolution of the shell structure in a specific mass-region.

The identification of isomeric states in the neutron-rich
nuclei with Z$\sim$28 and N$\sim$40-50 constitutes a field of a
great current interest. These states store key information about
the structural changes induced by increasing the neutron number
and especially by the filling of the unique parity $\nu g_{9/2}$
orbital. An experiment employing the fragmentation of a
$^{86}$Kr$^{34+}$ beam with an energy of 60.3 MeV/nucl. led to the
identification of thirteen new $\mu$s-isomers in the
neutron-rich nuclei from Sc ($Z$=21) to As ($Z$=33) \cite{Grz98}.
Spins and parities assignments were based on
the observed $\gamma$-decay pattern and comparisons with the systematics. Most of the
identified isomers were found to originate from the stretched $\nu
g_{9/2}^n$ configurations and decay to the lower-lying states via
$E2$ or $M2$ $\gamma$-transitions \cite{Grz98}.

In recent years, $\beta$-decaying isomers in the neutron-rich
nuclei around $^{68}$Ni have been extensively studied as well
\cite{Mue99,Pri99,Wei02,Jan1,Pau08}. Such isomers, arising from the large spin
difference between the opposite parity orbitals $\nu p_{1/2}$ and
$\nu g_{9/2}$, are expected to be found at low excitation energies in
the odd-odd and odd-N nuclei with N$\sim$40. Among these, the
odd-N Ni isotopes are of special interest since their low-energy
levels are expected to arise mainly from neutron single-particle
excitations whose investigation offers important information
concerning the core properties of $^{68}$Ni.

Low-lying states in the neutron-rich $^{67,69,71,73}$Ni were
identified in the $\beta$ decay of their Co isobars obtained in
proton-induced fission combined with resonant laser ionization
\cite{Mue99,Wei99} and in fragmentation reactions
\cite{Grz98,Pri99,Saw04,Raj07}. 1/2$^-$ and 9/2$^+$ spins and parities were proposed
for the ground-states of $^{67}$Ni and $^{69,71,73}$Ni,
respectively, based on the observed $\beta$-decay pattern and
shell-model calculations \cite{Mue99,Pri99,Wei99,Saw04,Raj07,Lis04}. In
$^{69}$Ni, the 1/2$^-$ state originating from $2p-1h$ excitations $\nu (p_{1/2}^{-1}g_{9/2}^2)$
across the N=40 subshell was found to be a long-lived isomer
($T_{1/2}$=3.5(5) s) decaying via a fast Gamow-Teller $\beta$ transition (log$ft=4.3(2)$)
to the first excited 3/2$^-$ state at 1298 keV in $^{69}$Cu
\cite{Fra98,Fra01,Mue99,Pri99}. The
observed weak branch to the 3/2$^-$ ground-state of $^{69}$Cu was
explained by invoking some mixing of the wave function, dominated
by the $\pi p_{3/2}^{+1} \nu g_{9/2}^{+0}$ component, with the $\pi
p_{3/2}^{+1}\nu p_{1/2}^{-2}g_{9/2}^{+2}$ configuration proposed for the
excited 3/2$^-$ level at 1298 keV \cite{Mue99,Pri99}. No feeding
of the isomer to the single-particle 1/2$^-$ state at 1096 keV in
$^{69}$Cu was observed, suggesting a rather pure $\pi p_{1/2}^{+1}\nu
g_{9/2}^{0}$ structure for this level.

Level schemes of the neutron-rich odd-A Ni isotopes beyond
$^{69}$Ni are still poorly known. In all odd-A Ni isotopes with masses from 71 to 77, the shell model predicts a spin and parity 9/2$^+$ for the
ground state and a low-lying 1/2$^-$ level dominated by the $\nu p_{1/2}$ neutron-hole configuration \cite{Lis04}. Sawicka {\it et al.} \cite{Saw04} reported four
$\gamma$ transitions in each of the $^{71,73}$Ni isotopes observed in the $\beta$ decay
of the $^{71,73}$Co isobars. Due to the poor statistics,
$\gamma$-$\gamma$ coincidences could not be constructed. Therefore, the
observed transitions were placed in a level scheme based on shell model predictions
\cite{Saw04}.

Most of the $\gamma$ rays reported by Sawicka {\it et al.} \cite{Saw04} were recently confirmed by the results of a decay-spectroscopy experiment performed at NSCL, MSU \cite{Raj07}. In that experiment, the $^{71,73}$Co isotopes were produced in the fragmentation of a $^{86}$Kr beam with an energy of 140 MeV/nucleon onto a thick $^{9}$Be target. The secondary fragments were implanted in a double-sided silicon strip detector surrounded by the NSCL Ge detector array SeGA used to detect the $\gamma$ rays. The good statistics obtained in that measurement allowed for the analysis of $\gamma$-$\gamma$ coincidences which provided the basis for the placement of the observed transitions in the decay schemes of $^{71,73}$Co given in Fig.~4 of Ref.~\cite{Raj07}. The two strong peaks observed at energies 566 and 774 keV did not show any coincidence relationship with each other nor with any of the other transitions observed in the $\gamma$ spectrum associated with the $\beta$ decay of $^{71}$Co. Based on a comparison with the systematics and shell-model predictions, both transitions were tentatively placed to feed the 1/2$^-$ level, suggested to be a $\beta$-decaying isomer \cite{Raj07}.

\begin{figure}
\includegraphics[height=6cm,width=9cm]{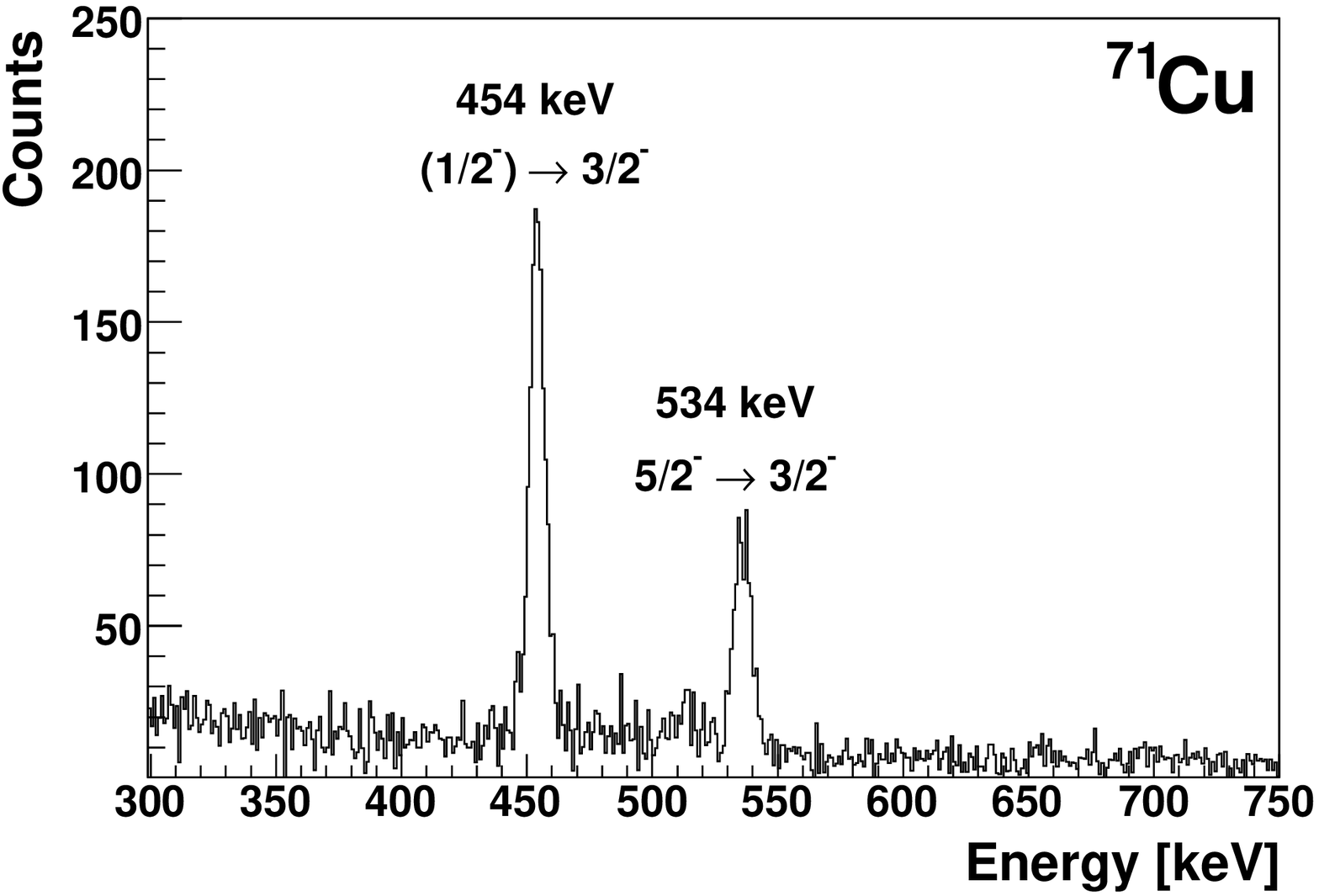}\\
\includegraphics[height=6cm,width=9cm]{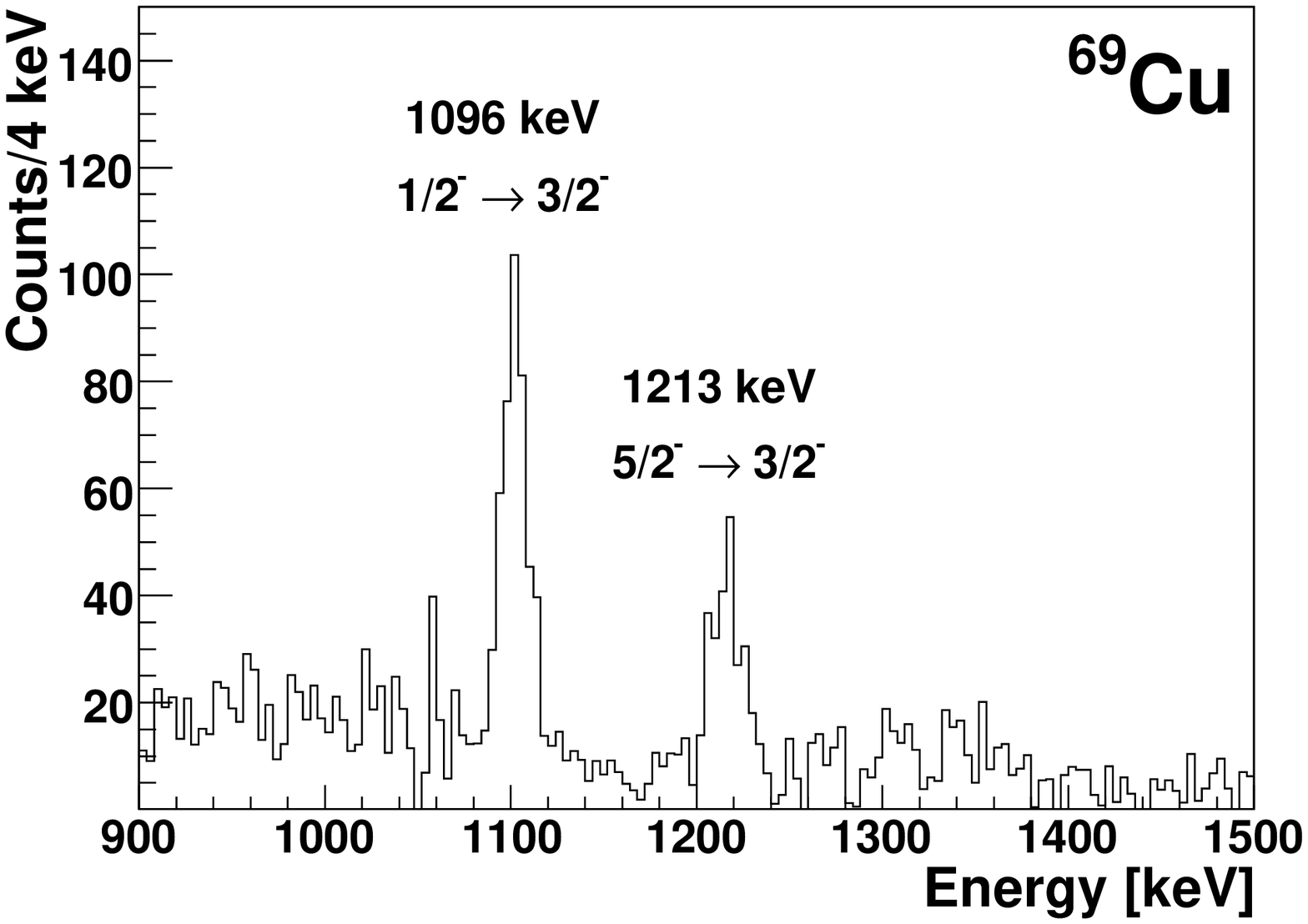}
\caption{\label{clx} Particle-$\gamma$-ray coincidence
spectra obtained after Coulomb excitation experiment with
radioactive beams of $^{71}$Cu (top) and $^{69}$Cu (bottom). The spectra are Doppler corrected for the mass of the projectile.}
\end{figure}

The present study was prompted by the observation of a new state
located at 454 keV populated in a recent Coulomb excitation
experiment with postaccelerated radioactive beam of $^{71}$Cu
produced at ISOLDE, CERN. The results of that measurement are
presented in Ref. \cite{Ste08}. The low beam-energy
($\sim$3 MeV/u) used in that experiment ensured that the
population of the excited states proceeds mainly via $E2$
excitations from the 3/2$^-$ ground-state, therefore only levels
with spins 1/2$^-$, 3/2$^-$, 5/2$^-$ and 7/2$^-$ were expected to
be populated. The top of Figure \ref{clx} shows the $\gamma$ spectrum after Coulomb
excitation obtained with the beam of $^{71}$Cu. The spectrum is Doppler
corrected for the mass of the projectile. The newly observed
$\gamma$ ray of 454 keV and the known 5/2$^-\rightarrow
3/2^-_{g.s.}$ transition of 534 keV \cite{Fra98,Fra01,Ish98} are clearly
visible in the spectrum.

\begin{figure*}
\includegraphics[height=8cm,width=14cm]{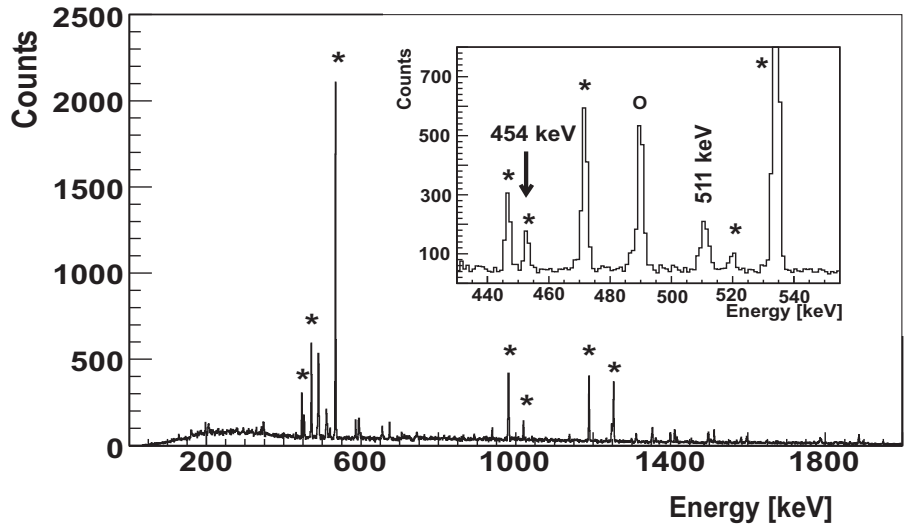}
\caption{\label{bg} Beta-gated $\gamma$ spectrum for mass $A$=71
obtained from the data set reported in \cite{Fra98,Fra01}. The $\gamma$ rays following the $\beta$ decay of $^{71}$Ni are marked with an asterisk. In the inset, the region
around 500 keV is enlarged, showing the 454-keV line. The
peak marked with a circle was assigned to the $\beta$ decay
of the daughter nucleus $^{71}$Cu.}
\end{figure*}

In the lighter Cu isotopes, an 1/2$^-$ state dominated by the $\pi
2p_{1/2}$ single-particle orbital was identified at low excitation
energies. Its energy is very close to that of the 5/2$^-$ level
found to contain a large component from the $\pi 1f_{5/2}$ orbital
\cite{Zei78}. The Coulomb excitation experiment mentioned above
included also a measurement with radioactive beam of $^{69}$Cu \cite{Ste08}. A
portion of the Doppler-corrected particle-$\gamma$ coincidence
spectrum obtained in that run is shown at the bottom of Figure
\ref{clx}. The two peaks present in the selected energy range were
identified as the transitions depopulating the first and second
excited states in $^{69}$Cu, namely the 1/2$^-$ and 5/2$^-$ levels
at 1096(6) \cite{Sel81} and 1213.5(1) keV \cite{Fra01}, respectively. Population
of the closely-lying 3/2$^-$ state at 1297.9(1) keV was not observed in the aforementioned  Coulomb excitation measurement.

As pointed above, the $E2$ excitation from the 3/2$^-$ ground-state of $^{71}$Cu constrains the spin of the newly identified state to values $I^\pi\leq$7/2$^-$. A spin 7/2$^-$ would imply a pure $E2$ character for the 7/2$^-\rightarrow$3/2$^-$ depopulating transition. The calculated Weisskopf estimate for the partial decay-lifetime indicates that an $E2$ transition of 454 keV will proceed in $\sim$2 ns, more than three orders of magnitude slower than an $M1$ transition. The observation of a Doppler broadened 454-keV peak in our Coulomb excitation spectrum suggests a half-life in the picosecond range for the emitting level and therefore an $M1$ character for the depopulating $\gamma$ ray. This restricts the spin of the 454 keV level to values $I^\pi\leq$5/2$^-$. In Ref.~\cite{Ste08}, spin and parity (1/2$^-$) were assigned to the newly observed state at 454 keV, based on the systematics of the lighter Cu isotopes and comparison with the
Coulomb excitation spectrum with beam of $^{69}$Cu. Such a spin assignment is also in agreement with the shell-model and particle-core coupling calculations \cite{Smi04,Fra01,Gra99,Oro00}. It is worth mentioning that in $^{71}$Cu, the second 3/2$^-$ state is still unknown. Shell-model and
particle-core calculations predict this state around 1900, 1662, and 1100 keV, respectively \cite{Smi04,Fra01,Oro00}.


It is also worthwhile to mention that isomeric 1/2$^-$ states originating from $2p-1h$ excitations across $Z$=40 have been observed in the valence partner of $^{71}$Ni, $^{93}_{43}$Tc$_{50}$ \cite{nndc}.
In fact, the 1/2$^-$ isomer was found to be the first excited state in the $N$=50 isotones from Nb ($Z$=41) to Rh ($Z$=45) with a half-life ranging from 60.9 days to 1.96 minutes, respectively
\cite{nndc}. In $^{93}$Tc, however, the $\beta$-decay from the 1/2$^-$ isomer ($T_{1/2}$=43.5 min) was found to compete with a $M4$ transition to the 9/2$^+$ ground-state \cite{nndc}.

This paper focuses on the possible decay modes of the 1/2$^-$
state in $^{71}$Ni. We analyze and discuss the experimental evidence indicating that this state is a $\beta$-decaying isomer feeding the newly observed (1/2$^-$) state at 454 keV in $^{71}$Cu \cite{Ste08}. The data sets used in the present study were obtained
in two different $\beta$-decay experiments performed at the LISOL
facility, Louvain-la-Neuve. In the first measurement, the
$\beta$-decay study of $^{71}$Ni was used as a means to
investigate the low-lying level scheme of $^{71}$Cu
\cite{Fra98,Fra01}. The second experiment was dedicated to the
identification of the energy levels in $^{71}$Ni populated in the
decay of the $^{71}$Co isobar.

\section{EXPERIMENTAL DETAILS}

The $^{71}$Ni and $^{71}$Co beams were produced in two separate
measurements at the LISOL facility by colliding a 30-MeV proton
beam with two thin $^{238}$U foils mounted inside a gas cell \cite{Kud96}. The
cell was filled with 500 mbar of Argon gas. The radioactive Ni and Co
atoms were resonantly photoionized, mass separated and implanted
in a movable tape surrounded by $\beta$ and $\gamma$ detectors
arranged in a close geometry. Table \ref{lln} gives a summary of
the experimental conditions in both measurements.

\begin{table*}
\caption{\label{lln}Half lives of the mother nuclei, specific
implantation-decay cycles, measuring times with and without laser
radiation, beam intensities and productions rates for $^{71}$Ni
and $^{71}$Co. During experiment I, the lasers were tuned on nickel, while during experiment II they were tuned on cobalt.}
\begin{ruledtabular}
\begin{tabular}{cccccccc}
Exp.&Nucleus&$T_{1/2}$&Cycle&Laser&Laser&I$_{beam}$&Yield\\
no&&(s)&(impl./decay)&ON&OFF&($\mu$A)&(ions/$\mu$C)\\
\hline
I&$^{71}$Ni&2.56(3)\footnotemark[1]&6 s/10 s&35 h 09 min&-&6.1&3.0(6)\\
II&$^{71}$Co&0.079(5)\footnotemark[2]&0.6 s/1 s&16 h 02 min&12 h 57 min&6.7&0.032(8)\\
\end{tabular}
\end{ruledtabular}
\footnotetext[1]{Ref. \cite{Fra01}.} \footnotetext[2] {Ref.
\cite{Saw04}.}
\end{table*}

In the first experiment, $^{71}$Ni was implanted in a Mylar tape
surrounded by two HPGe detectors positioned in the horizontal
plane and at 90$^\circ$ and 270$^\circ$ with respect to the beam
axis. The relative efficiency of the detectors reached 70$\%$ and
75$\%$, respectively. The emitted beta particles were recorded in a plastic $\Delta E$ scintillator located between the two Ge detectors, in forward direction. A detailed description of the experimental
setup can be found in \cite{Fra01}.


In the second experiment, the $\beta$ decay of $^{71}$Co was
observed by means of four plastic $\Delta E$ detectors while the
emitted $\gamma$ rays were recorded with three HPGe detectors of
70$\%$, 75$\%$ and 90$\%$ relative efficiency located at
90$^\circ$, 0$^\circ$ and 270$^\circ$, respectively, with
respect to the beam axis. Measurements with and without laser
radiation were performed in order to disentangle the $\gamma$ rays
emitted by the nuclei of interest from the non-resonant
transitions.

\section{EXPERIMENTAL RESULTS}
\subsection{\label{ni} $\beta$ decay of $^{71}$Ni}

The $\beta$-gated $\gamma$ spectrum obtained in the experiment
with $^{71}$Ni beam is presented in Fig. \ref{bg}. In the inset,
the region around 500 keV is enlarged, showing the transition of
454 keV. Gamma rays attributed to the decay of $^{71}$Ni (see Ref.
\cite{Fra01} for details) are marked with an asterisk. Open
circles label the transitions arising from the $^{71}$Cu decay.

In Ref. \cite{Fra01}, the 454-keV transition was found not to be in coincidence with any of the
$\gamma$ rays attributed to $^{71}$Cu, therefore it was not
further discussed in that paper. The proposed spin value (1/2$^-$)
for the 454 keV level out of the Coulomb excitation study rules out the possibility of a direct $\beta$ branch from the 9/2$^+$ ground state of $^{71}$Ni. Furthermore, the lack of $\gamma$-ray coincidences in the $\beta$ decay study of \cite{Fra01} also excludes indirect feeding from the  ground state of $^{71}$Ni, see further. Therefore, the alternative scenario that this state is fed by a 1/2$^-$ $\beta$-decaying isomer in $^{71}$Ni is investigated.

\begin{figure}[h!]
\includegraphics[height=6cm,width=9cm]{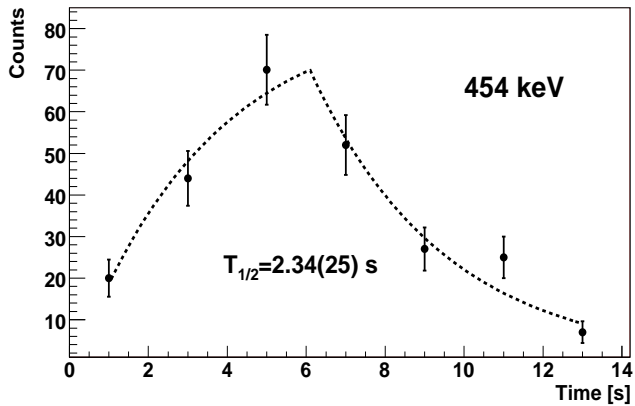}
\caption{\label{time} Time evolution of the intensity of the 454
$\gamma$-ray fitted with a single exponential function yielding to
a half-life of $T_{1/2}$=2.3(3) s.}
\end{figure}

The time evolution of the $\gamma$ intensity of the 454-keV
transition is shown in Fig. \ref{time}. The data were fitted with a
single exponential. The poor statistics forced us to take bins
of 2 seconds each resulting in three and five data points for the
implantation and decay periods, respectively (see Table
\ref{lln}). The last two seconds of the decay period were excluded
due to the very low number of counts observed in the peak. A value
of $T_{1/2}$=2.34(25) s was extracted from the fit.

\begin{figure}
\includegraphics[height=9cm,width=9cm]{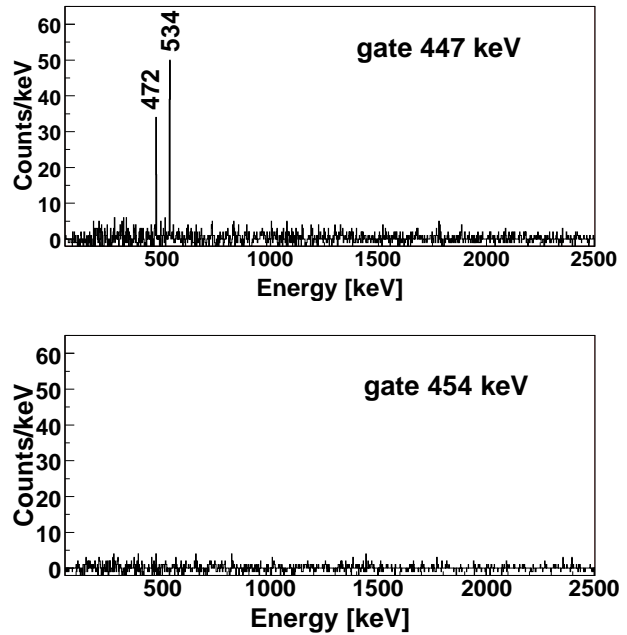}
\caption{\label{coinc} Background-subtracted
$\gamma$-$\gamma$ coincidence spectra gated on the 447 keV transition (top) and 454 keV $\gamma$-ray (bottom).}
\end{figure}

Figure \ref{coinc} compares the $\gamma$-$\gamma$ spectra gated with the 447-keV
transition (top) and the 454-keV line (bottom). The 447-keV transition deexcites the level at 981 keV which is populated both directly in the $\beta$-decay of the 9/2$^+$ ground-state of $^{71}$Ni and from feeding from higher-lying states in $^{71}$Cu \cite{Fra01}. As can be seen from the inset of Fig. \ref{bg}, the 447-keV line is twice stronger than the peak at 454 keV. The observation of the 472- and 534-keV transitions in the spectrum coincident with the 447 keV $\gamma$ ray provided the basis for its placement in the level scheme of $^{71}$Cu as shown in Fig. 7 of Ref. \cite{Fra01}. In the spectrum gated with the 454-keV $\gamma$ ray shown in the bottom of Fig. \ref{coinc}, no clear peak can be distinguished from the background, indicating  that $\gamma$ feeding from higher-lying states has a negligible contribution to the population of the 454-keV level. The observed background is due to true coincidences with $\beta$ particles interacting with the Ge detectors. The non-observation of any coincident $\gamma$ ray supports the scenario that this state is directly fed by a 1/2$^-$ isomer in $^{71}$Ni.

\begin{figure*}
\includegraphics[width=\linewidth]{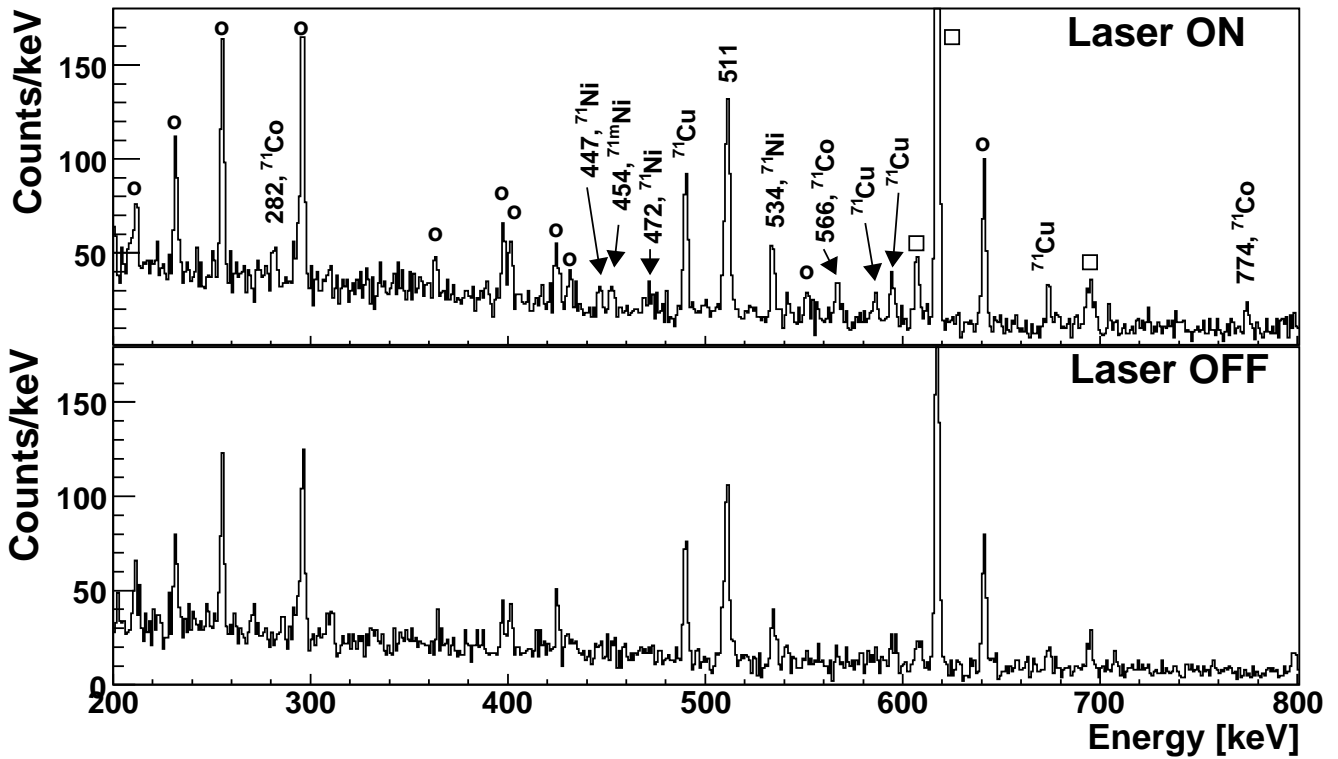}
\caption{\label{71co} Beta-gated $\gamma$ spectrum for A=71 when
laser are on (top) Co resonance and lasers are off (bottom). Transitions
belonging to the decay of $^{71}$Co (see text), $^{71}$Ni and $^{71}$Cu are marked on
the figure. Contaminant $\gamma$ lines are marked with open circles. The open squares mark $\gamma$ lines belonging to the decay of $^{112}$Ag, which originates from $^{112}$Rh isotopes implanted next to the tape during the optimization of the laser-ion source.}
\end{figure*}

From the observed number of counts in the 454-keV peak and taking into account the absolute $\gamma$-ray branching from the (1/2$^-$) $\beta$-decaying isomer (see section \ref{s71co}) we
extract a production rate of $0.2 (1)$ at/$\mu$C of the $1/2^{-}$ isomer in experiment I. With a ground state yield of $3.0(6)$ at/$\mu$C, see Table \ref{lln}, this results in an isomeric ratio of $7 (4) \%$. Within the same experimental conditions, a lower limit of 0.74 at/$\mu$C was reported for the production of the (1/2$^-$) isomer in $^{69}$Ni, which was found to represent nearly 20$\%$ from the total production rate of $^{69}$Ni \cite{Fra01}.

\subsection{\label{s71co}$\beta$ decay of $^{71}$Co}

Figure \ref{71co} shows part of the $\beta$-gated $\gamma$-spectra
observed in the measurement with the lasers tuned to ionize Co
(top) and without laser radiation (bottom). Transitions belonging
to the $A$=71 decay chain are labeled on the figure. The
contaminant lines, represented with open circles, were found to be
emitted by the $^{142}$La or $^{102}$Nb fission products. $^{142}$La as well as $^{102}$Nb in the form of a molecule with $^{40}$Ar could reach the detection setup in a 2$^+$ charge state and therefore with the same $A/q$ ratio as the ions of interest. Another source of contamination found to give a non-resonant signal in the $\beta$-gated $\gamma$ spectra shown in Fig. \ref{71co} was $^{112}$Ag, produced in the decay of $^{112}$Rh which was implanted next to the tape during the optimization of the laser-ion source.

With the lasers tuned on the Co resonance, the transitions of 282, 566
and 774 keV, assigned to the decay of $^{71}$Co in Refs. \cite{Saw04,Raj07}, are clearly visible in Fig. \ref{71co}(top).

The lasers-off spectrum (Fig. \ref{71co}, bottom) shows the presence of the 534-keV line from the non-resonant production of $^{71}$Ni. The upper limits of the $\gamma$ intensities of the other lines from the decay of $^{71}$Ni are consistent with Ref. \cite{Fra01}. However, in the Co on resonance spectrum (Fig. \ref{71co}, top), an excess of 35(11) counts in the 454-keV line was observed when using the intensity ratio  $I_\gamma(534)/I_\gamma(454)$ from Ref. \cite{Fra01} (see Fig. \ref{bg}).
This confirms that the 454-keV line is populated in the decay of the newly identified (1/2$^-$) isomer of $^{71}$Ni, which in turn is fed by the $\beta$-decay of $^{71}$Co \cite{Raj07}. A comparison of the intensity of the 534-keV line in the top and bottom spectra of Fig. \ref{71co} indicates that in the Co on resonance spectrum the observed intensity mainly stems from non-resonant production of $^{71}$Ni. Thus, our data do not allow to confirm the $\sim$10$\%$ indirect feeding of the $^{71}$Ni ground state in the $^{71}$Co decay \cite{Raj07}. However, using the $^{71}$Co decay scheme of Ref.~\cite{Raj07}, we can deduce the absolute 454-keV $\gamma$-ray intensity.

From the intensities of the 566-, 774-, and 454-keV lines observed in the spectrum displayed in Fig.~\ref{71co}, we determined an absolute $\gamma$-ray branching of $40(15) \%$ to the 454-keV transition. In Fig.~\ref{71co}, the absolute $\gamma$-ray branching is taken as the direct $\beta$ branching to the $454$-keV level and the remaining intensity of $60(15) \%$ is attributed to the direct feeding of the ground state of $^{71}$Cu. Because weaker $\gamma$ transitions to both the $454$-keV level and ground state might have been missed, the $\beta$ branching values should be considered as upper limits. The value extracted for the feeding to the excited state corresponds to a log$ft$ of 5.4(2), assuming that the (1/2$^-$) isomer in $^{71}$Ni is indeed located at 499 keV \cite{Raj07}.

\section{DISCUSSION}

\begin{figure*}
\includegraphics[width=0.8\linewidth]{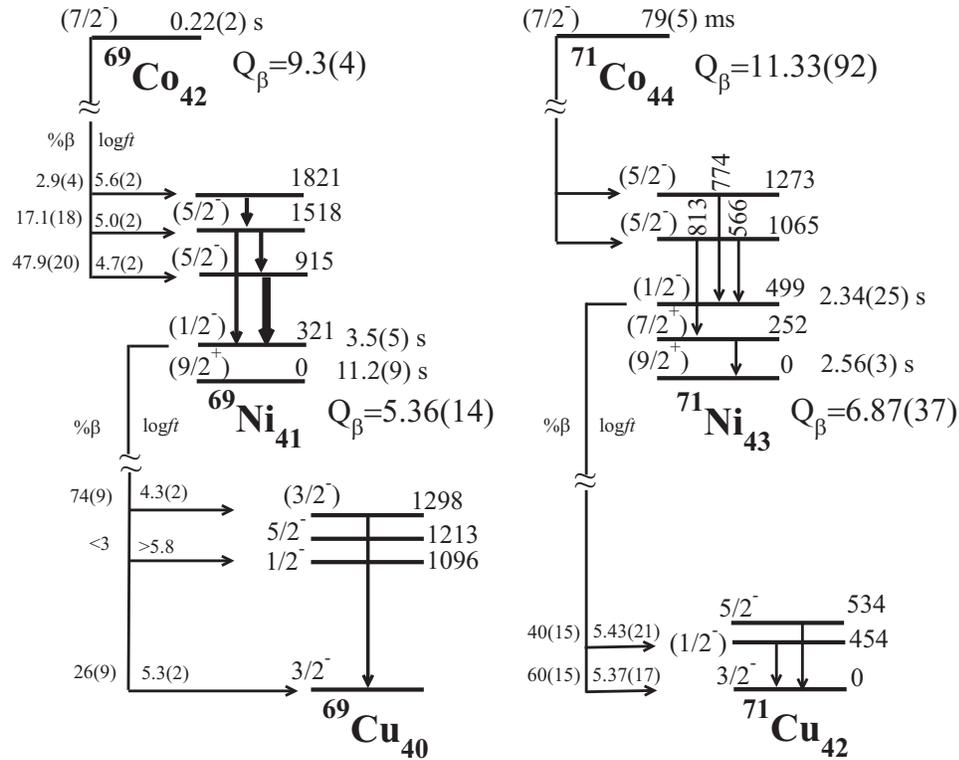}
\caption{\label{schemes} Observed $\beta$-decay chain $^{69}$Co $\rightarrow$ $^{69}$Ni $\rightarrow$ $^{69}$Cu \cite{Mue99,Fra01} and  $^{71}$Co $\rightarrow$ $^{71}$Ni $\rightarrow$ $^{71}$Cu as proposed in Ref. \cite{Raj07} and present work. $Q_\beta$ values are given in MeV.}
\end{figure*}

Let us now discuss the implications of the present findings on the
evolution of the nuclear structure in this mass-region.
Figure \ref{schemes} shows the comparison between the $\beta$-decay chains $^{69}$Co $\rightarrow$ $^{69}$Ni $\rightarrow$ $^{69}$Cu \cite{Mue99,Fra01} and $^{71}$Co $\rightarrow$ $^{71}$Ni $\rightarrow$ $^{71}$Cu \cite{Fra01,Raj07}. Both $^{69,71}$Co isotopes are assumed to have a 7/2$^-$ ground-state dominated by the $\pi f_{7/2}^{-1}$ proton-hole configuration for which the major decay path is the Gamow-Teller decay of a $f_{5/2}$ core neutron to fill the  $f_{7/2}$ proton orbital. In $^{69}$Ni, the strong $\beta$-decay branches from $^{69}$Co observed to the levels at 915 and 1518 keV suggested dominant $\nu f_{5/2}^{-1}\otimes 0^+(^{70}$Ni) and $\nu f_{5/2}^{-1}\otimes 2^+(^{70}$Ni) configurations, respectively, although the latter is considerably mixed up with a $\nu p_{1/2}^{-1}$ $\otimes$ $2^+(^{70}$Ni) component \cite{Mue99}. As discussed in Ref. \cite{Raj07}, the expected strong Gamow-Teller $\beta$-decay branch from the 7/2$^-$ ground-state of $^{71}$Co restricts the spins and parities of the excited states populated in the $^{71}$Ni daughter nucleus to 9/2$^-$, 7/2$^-$ and 5/2$^-$. Based on shell-model predictions and the observed systematics, spin and parity 5/2$^-$ were assigned to the levels at 1065 and 1273 keV in $^{71}$Ni, see Ref. \cite{Raj07} and Fig. \ref{schemes}.

In both $^{69,71}$Ni nuclei, the 5/2$^-$ states receiving the main $\beta$-feeding are assumed to decay via $E2$ transitions towards the (1/2$^-$) isomer. In $^{71}$Ni, however, shell-model predicts that the first excited level is 7/2$^+$ \cite{Lis04}. The presence of such state below the (1/2$^-$) level reduces the spin difference between the 5/2$^-$ levels populated in $\beta$-decay and the 9/2$^+$ ground-state and increases the probability for $\gamma$-decay 5/2$^-\rightarrow 7/2^+ \rightarrow 9/2^+_{g.s.}$, by-passing the (1/2$^-$) isomer. Based on the analysis of $\gamma$-$\gamma$ coincidences, the observed 813-252 keV cascade was assigned to this spin-sequence in Ref. \cite{Raj07}.


The $\beta$-decay of the (1/2$^-$) isomer in $^{69}$Ni was found to populate essentially the state at 1298 keV in $^{69}$Cu, see Fig. \ref{schemes}. From the comparison of the $\gamma$-intensity feeding into the 321-keV level in $^{69}$Ni with the intensity of the 1298-keV transition in $^{69}$Cu, a $\beta$-branching of 74(9)$\%$ was determined in Ref. \cite{Mue99} for the level at 1298 keV. This branching corresponds to a log$ft$ value of 4.3(2), see Ref. \cite{Mue99} and Fig. \ref{schemes}. Spin and parity 3/2$^-$ were assigned to the 1298-keV state, viewed as the $p_{3/2}$ proton coupled to the $2p-2h$,  0$^+$ state at 1770 keV in the $^{68}$Ni core \cite{Mue99}. Such configuration implies very low collectivity for the 3/2$^-$state, in agreement with its non-observation in the Coulomb excitation spectrum shown in Fig. \ref{clx}. In contrast to the 3/2$^-$ single-particle level at 1298 keV in $^{69}$Cu, the state at 454 keV in $^{71}$Cu was found to exhibit large collectivity $(B(E2;1/2^- \rightarrow 3/2^-_{g.s.})$=20.4(22) W.u. as determined in Ref. \cite{Ste08}). By relating the number of counts in the peaks at 1096 and 1213 keV observed in the bottom spectrum of Fig. \ref{clx} with the corresponding $B(E2)$ values reported in Ref. \cite{Ste08}, an upper limit of 1.4 W.u. can be extracted for the $B(E2)$ value for the 1298-keV transition. Thus, the decay of (1/2$^-$) isomer in $^{69,71}$Ni populates states with very different character in the daughter nuclei.

In $^{71}$Ni, however, our evidence shows that the $\beta$-decaying isomer feeds mainly the proposed (1/2$^-$) state at 454 keV in $^{71}$Cu. As discussed in Ref. \cite{Ste08}, the large $B(E2)$ value measured for the 454 keV transition excludes a single-particle character of $\pi p_{1/2}$ type for the 454 keV level. The increased collectivity indicates significant deformation setting in with increasing the number of $g_{9/2}$ neutrons, as also suggested by the results of recent Coulomb excitation experiments with radioactive
beams of $^{70}$Ni \cite{Per06}. The onset of collectivity was associated with the quenching of both $Z$=28 and $N$=40 gaps through the combined effect of the attraction and repulsion between the $fp$ protons and $g_{9/2}$ neutrons \cite{Per06,Ste07,Ots05}.  Thus, the observed $\beta$-decay branch from the (1/2$^-$) isomer in $^{71}$Ni to the 454-keV level in $^{71}$Cu  can be explained by assuming that the odd proton occupies the $K$=1/2 downsloping orbit of the $p_{3/2}$ orbital, on the prolate side, while the neutron part of the wave function is, depending on deformation, dominated by $\nu p_{1/2}^{+2}g_{9/2}^{+2}$ or $\nu p_{1/2}^{-2}g_{9/2}^{+4}$ configurations. This indicates that in both $^{69,71}$Ni isotopes, the $\beta$-decay of the (1/2$^-$) isomer proceeds via a fast Gamow-Teller transition but in the case of $^{71}$Ni the spin of the  final state in the daughter nucleus is changed by the deformation. Interesting to note is that a similar deformed $\pi 1/2^-[321]$ has been observed in $^{67}$Co, stemming from a $\pi \text{(1p-2h)}$ proton excitation across $Z$=28 \cite{Pau08}.

\section{Conclusions}

In this paper, the results of the investigation of the decay of the proposed (1/2$^-$) $\beta$-decaying isomer in $^{71}$Ni are presented and discussed. The key observable for this study is the newly observed level at 454 keV in $^{71}$Cu reported recently in Ref.~\cite{Ste08} and for which the comparison with the systematics and model calculations predict a spin and parity of 1/2$^-$. The experimental evidence discussed here combines the results of the Coulomb excitation measurement with radioactive beams \cite{Ste08} with the results of two decay experiments aiming to the investigation of the $\beta$-decay of $^{71}$Co and $^{71}$Ni. The analysis of the $\beta$-decay of $^{71}$Ni indicates that the 454-keV state observed in $^{71}$Cu is fed by the (1/2$^-$) $\beta$-decaying isomer in $^{71}$Ni for which a half-life of $T_{1/2}$=2.34(25) s was determined in the present work.

The large $B(E2)$ value measured in Ref. \cite{Ste08} for the 454-keV transition depopulating the (1/2$^-$) state in $^{71}$Cu indicated a deformed structure for this level. This indicates that in both $^{69,71}$Ni isotopes the main $\beta$-decay branch of the (1/2$^-$) isomer goes to the level dominated by the $\pi p_{3/2} \nu (p_{1/2}^{-2}g_{9/2}^2)$ configuration in the daughter nuclei. In $^{71}$Cu, however, due to deformation, the nuclear properties of the level receiving the main $\beta$-feeding are dictated by the $K$=1/2 downsloping orbit of the $\pi p_{3/2}$ orbital.

We gratefully thank J.~Gentens and P.~Van den Bergh for running the
LISOL separator. This work was supported by the European Commission within the Sixth Framework Programme through
I3-EURONS (Contract RII3-CT-2004-506065), BriX/IUAP P6/23, FWO-Vlaanderen (Belgium), GOA 2004/03 and US Department of Energy.

\bibliography{apssamp}

\end{document}